\documentclass[aps,prd,a4paper,onecolumn,amsmath,showpacs,superscriptaddress,nofootinbib,preprintnumbers,notitlepage]{revtex4-1}
 
\usepackage{verbatim}
\usepackage[T1]{fontenc}
\usepackage[utf8]{inputenc}
\usepackage[american]{babel}
\usepackage{epsfig}
\usepackage{array}
\usepackage{graphicx,subcaption,caption}
\usepackage{pgfplots}
\usepackage{booktabs}
\usepackage{multirow}
\usepackage{dcolumn}
\usepackage{amsmath}
\usepackage{mathtools}
\usepackage{amsfonts}
\usepackage{amssymb}
\usepackage{epstopdf}
\usepackage{bm}
\usepackage{siunitx}
\usepackage{braket}
\usepackage{enumitem}
\usepackage{soul}
\usepackage{color}
\usepackage{transparent}
\usepackage{pifont}
\usepackage[font={small}]{caption}

\definecolor{navyblue}{rgb}{0.0, 0.0, 0.5}
\definecolor{royalblue}{rgb}{0.25, 0.41, 0.88}
\definecolor{cadmiumgreen}{rgb}{0.0, 0.42, 0.24}
\definecolor{blue-violet}{rgb}{0.54, 0.17, 0.89}
\definecolor{darkviolet}{rgb}{0.58, 0.0, 0.83}
\definecolor{orange(colorwheel)}{rgb}{1.0, 0.5, 0.0}

\usepackage{hyperref}
\hypersetup{
    colorlinks=true, 
    linkcolor=royalblue, 
    citecolor=magenta}

\newcommand\be{\begin{equation}}
\newcommand\ee{\end{equation}}
\newcommand\bea{\begin{eqnarray}}
\newcommand\eea{\end{eqnarray}}

\newcommand\ie{{\it i.e.}~}

\usepackage{booktabs}
\usepackage{multirow}
\usepackage{dcolumn}
\usepackage{colortbl}

\definecolor{magenta(process)}{rgb}{1.0, 0.0, 0.56}

\definecolor{darkspringgreen}{rgb}{0.09, 0.45, 0.27}

\definecolor{royalblue(web)}{rgb}{0.25, 0.41, 0.88}

\begin{document}

\title{Bekenstein bound on black hole entropy in non-Gaussian statistics}

\author{Mehdi Shokri}
\email{mehdishokriphysics@gmail.com}

\affiliation{Canadian Quantum Research Center, 204-3002 32 Ave, Vernon, BC V1T 2L7, Canada}
\affiliation{School of Physics, Damghan University, P. O. Box 3671641167, Damghan, Iran}

\preprint{}
\begin{abstract}
The Bekenstein bound, inspired by the physics of black holes, is introduced to constrain the entropy growth of a physical system down to the quantum level in the context of a generalized second law of thermodynamics. We first show that the standard Bekenstein bound is violated when the entropy of a Schwarzschild black hole is described in non-Gaussian statistics Barrow, Tsallis, and Kaniadakis due to the presence of the related indices $\Delta$, $q$ and $\kappa$, respectively. Then, by adding the GUP effects into the Bekenstein bound, we find that the generalized bound is satisfied in the context of the mentioned entropies through a possible connection between the entropies indices and the GUP parameter $\beta$. 
\end{abstract}
\maketitle
\section{Introduction}

Black holes provide a framework to pursue an essential link between gravitational physics and thermodynamics \cite{Bardeen:1973gs}. Moreover, further studies by Bekenstein and Hawking confirmed that quantum physics plays a noteworthy role in analyzing the thermodynamic quantities of black holes \cite{Bekenstein:1973ur,Hawking:1975vcx}. In 1973, Bekenstein claimed that the black hole entropy, as a thermodynamic quantity, directly connects to its horizon area as a geometrical object \cite{Bekenstein:1972tm,Bekenstein:1973ur}. Later, Hawking approved Bekenstein's conjecture by introducing the Bekenstein–Hawking (BH) formula $S = A/4$. From a purely classical thermodynamics perspective, black holes act like a perfect absorber, without any thermal emission, with an absolute zero temperature. However, Hawking discovered black holes emit thermal radiation in the presence of quantum effects \cite{Hawking:1974rv,Hawking:1975vcx}. Hence, black holes can be understood in terms of the generalized laws of thermodynamics with quantum effects \cite{Bekenstein:1972tm,Bekenstein:1973ur,Hawking:1974rv,Hod:2015iig}. In principle, the mentioned quantum effects allow us to present an interpretation of the laws of black hole mechanics as physically corresponding to the ordinary laws of thermodynamics \cite{Wald:1999vt, Page:2004xp, Carlip:2014pma,Almheiri:2020cfm}. This point can be seen when we deal with the universal
bounds on the entropy of black hole systems in the context of the generalized second law of thermodynamics, telling us the entropy of a closed physical system will grow until it reaches its maximum value. The first attempt to limit the process of entropy growth of a physical system was made by Bekenstein using heuristic arguments of black holes in order to find an upper bound for the entropy of macroscopic bodies in terms of the total energy and radius of the smallest sphere that encloses the object \cite{Bekenstein:1980jp}. Although the Bekenstein bound conjecture was obtained by studying the physics of black holes, it does not contain Newton's constant $G$. Therefore, the Bekenstein bound is not restricted only to gravitational scenarios. Despite the quantum proofs of Bekenstein's discovery \cite{Bekenstein:1981zz,Bekenstein:1982zv,Bekenstein:2000ai,Schiffer:1989et,Bekenstein:1990du,Casini:2008cr}, several counterexamples point out its possible shortcomings \cite{Unruh:1982ic,Unwin:1982pg,Page:1982fj,Bekenstein:1983iq,Unruh:1983ir,Pelath:1999xt}. Further studies by 't Hooft and Susskind showed that the Bekenstein bound can not be applied to thermodynamic systems involving strong gravity \cite{tHooft:1993dmi,Susskind:1994vu}. They claimed that the entropy of our gravitational system should be bounded by area in the context of the holographic principle. Subsequently, Bousso argued that such a spacelike entropy bound is violated more broadly in many dynamical and cosmological situations. He suggested the covariant entropy bound or Bousso entropy bound by introducing a "lightsheet", defined as a region traced out by non-expanding light-rays emitted orthogonally from an arbitrary surface \cite{Bousso:1999xy}. A derivation of the Bousso bound was given by Flanagan, Marolf, and Wald by considering a condition on a null hypersurface \cite{Flanagan:1999jp}. 

In addition to the BH statistics, we might work with non-Gaussian statistics by taking into account some corrections to the BH entropy coming from non-extensivity of the horizon degrees of freedom (DOF) or quantum gravitational effects on the black hole surface area. These statistics provide a natural framework to study semiclassical effects in the context of the generalized uncertainty principle (GUP). For instance, Tsallis proposed a class of non-Gaussian entropies by considering some non-extensive effects to the Boltzmann-Gibbs statistics \cite{Tsallis:1987eu}, which is parameterized by index $q$. The non-extensivity parameter $q$ refers to super-extensivity and sub-extensivity when $q<1$ and $q>1$, respectively \cite{Tsallis:2002tp}. In fact, rare events occur for $0\leq q<1$ and frequent events occur for $q>1$ \cite{Tsallis:1998as,NIVEN2004444}, pointing to the stretching or compressing of the entropy curve to lower or higher maximum entropy positions. The consequences of these non-extensivity effects on the thermodynamic parameters have been discussed in \cite{Abe:2001ss}. In the context of the "gravity-thermodynamics" conjecture, a broad range of gravitational and cosmological scenarios have been studied in the presence of non-extensivity. Such as dark energy (DE) \cite{Barboza:2014yfe,Nunes:2015xsa,Zadeh:2018poj,Ghaffari:2018wks,Tavayef:2018xwx, SayahianJahromi:2018irq, Saridakis:2018unr,Sheykhi:2018dpn, Lymperis:2018iuz, Sheykhi:2019bsh,Huang:2019hex,Aditya:2019bbk,DAgostino:2019wko,Mamon:2020wnh,Mohammadi:2021wde,Dheepika:2021fqv,Nojiri:2022dkr,Nojiri:2022aof,Nojiri:2022dkr}, cosmic inflation  \cite{Keskin:2023ngx,Odintsov:2023vpj,Teimoori:2023hpv}, black holes \cite{deOliveira:2005jy,Tsallis:2012js,Komatsu:2013qia,Mejrhit:2020dpo,Abreu:2020wbz,Nojiri:2021czz,Nojiri:2023ikl}, gravitational waves \cite{Jizba:2024klq}, the cosmic microwave background (CMB) \cite{Bernui:2005hq,Bernui:2007wj} and modified Newtonian dynamics (MOND) \cite{Abreu:2014dna,Abreu:2018pua}. A relativistic generalization of the Boltzmann-Gibbs statistics was introduced by Kaniadakis \cite{Kana}, which naturally emerges from the special relativity perspective \cite{Kaniadakis:2002zz}. In $\kappa$-statistics, we deal with a non-exponential distribution function parameterized by the index $-1<\kappa<1$. See its applications in holographic dark energy (HDE) \cite{Drepanou:2021jiv,Hernandez-Almada:2021aiw,P:2022amn}, cosmic inflation  \cite{Nojiri:2022dkr,Odintsov:2023vpj,Lambiase:2023ryq}, black holes \cite{Abreu:2021avp,Abreu:2021kwu, Cimidiker:2023kle} and other cosmological situations \cite{Luciano:2022eio}. Another class of non-Gaussian statistics deals with logarithmic and power-law correction terms due to quantum gravitational effects in the physics of black holes. Considering self-gravitation and backreaction effects to the metric of the black hole navigates us to the BH logarithmic-corrected statistics described by two dimensionless and model-dependent constants $\alpha$ and $\beta$. For gravitational and cosmological applications of the logarithmic-corrected version of the BH entropy, see Refs.\cite{Jamil:2010xq,Cai:2010zw,Abreu:2020wbz,Abreu:2020dyu}. Recently, Barrow \cite{Barrow:2020tzx} has proposed a new type of statistics by considering a fractal structure, coming from quantum gravitational effects, on the surface of the Schwarzschild black hole tending to infinity with a finite volume. The Barrow entropy is parameterized by the index $0\leq\Delta\leq1$, which depicts the corresponding quantum gravity effects. Note that the quantum gravity effects considered in the Barrow entropy are different than the assumed quantum corrections in the context of the logarithmic-corrected entropy \cite{Kaul:2000kf}, however, it
resembles Tsallis entropy \cite{Tsallis:1987eu}, nevertheless the involved foundations and physical principles are
completely different. Also, the discussed fractal behavior does not arise from specific quantum gravity calculations, but from general simple physical principles, which adds to its plausibility and hence it
is valid as a first approach to the subject \cite{Barrow:2020tzx}. Besides phenomenological supports, the Barrow entropy is secured by some important theoretical proofs. In \cite{Saridakis:2020zol}, the authors used the standard
holographic principle at a cosmological structure and Barrow entropy. In \cite{Saridakis:2020cqq}, the authors calculated
the entropy time-variation in a universe filled with matter and dark energy fluids, as well as
the corresponding quantity for the apparent horizon. They showed that the generalized second law of thermodynamics is valid in the context of the Barrow statistics due to a small deviation from the Bekenstein-Hawking entropy ($\Delta=0$). Also, in \cite{Saridakis:2020lrg}, the authors discussed modified cosmological scenarios using Barrow entropy. Moreover, Barrow’s entropy has been tested,
theoretically and observationally, by studying the gravitational and cosmological situations such as Barrow holographic dark energy (BHDE) models \cite{Dabrowski:2020atl,Huang:2021zgj,Nojiri:2021jxf,Luciano:2022viz,Chanda:2022tpk}, inflation theory \cite{Maity:2022gdy,Luciano:2023roh,Saha:2024dhr} and black holes physics \cite{Abreu:2020dyu,Abreu:2020wbz,Abreu:2021kwu,Abreu:2022pil, Jawad:2022lww,Wang:2022hun,Abreu:2022pil,DiGennaro:2022grw,Nojiri:2022sfd}.

As we mentioned above, universal entropy bounds, which include Heisenberg uncertainty principle (HUP) effects, limit the process of the entropy growth for physical systems down to the quantum level. Hence, it is worthy to know how entropy bounds behave in the context of non-Gaussian entropies, which are affected by GUP effects. In the present manuscript, we attempt to clarify this issue by studying the Bekenstein bound for black hole systems represented by well-known non-Gaussian entropies Barrow, Tsallis, and Kaniadakis. All above information motivates us to arrange the paper as follows. In section \ref{s2}, we study HUP and GUP versions of the Bekenstein bound for thermal systems like black holes in the context of Barrow, Tsallis, and Kaniadakis statistics. 
We draw conclusions in section \ref{s3}. 

\section{Bekenstein bound conjecture in non-Gaussian entropies}\label{s2}
In 1981 J. Bekenstein introduced an upper bound on the entropy of a confined quantum system by using heuristic arguments involving black holes. The Bekenstein bound on the entropy $S$ of any object of maximal radius $R$ and total energy $E$ is given by \cite{Bekenstein:1980jp}
\begin{equation}
S\leq\frac{2\pi k_B}{\hbar c}RE,
\label{a1}    
\end{equation}
where $k_B$ is the Boltzmann constant. Note that, at the classical level, when  $\hbar\rightarrow0$, we have $S\leq\infty$, pointing out that the entropy of a localized system is unbound from above. As we see, the bound (\ref{a1}) does not include Newton's constant $G$, so it can be used for non-gravitational phenomena in addition to gravitational systems. In \cite{Buoninfante:2020guu}, the authors showed that the Bekenstein bound in the presence of GUP effects takes two following configurations:
\begin{equation}
  S\leq\begin{cases}
    \frac{2\pi k_B}{\hbar c}RE\Big[1-\frac{\beta}{4}\Big(\frac{\ell_{p}}{R}\Big)^2\Big], & \text{for $\beta>0$},\\\\
     \frac{2\pi k_B}{\hbar c}RE\Big[1+\frac{|\beta|}{4}\Big(\frac{\ell_{p}}{R}\Big)^2\Big], & \text{for $\beta<0$},\\
  \end{cases}
  \label{a2}
\end{equation}
which are obtained in the limit of $\beta \ell_p/R\ll1$. Clearly, the generalized bound (\ref{a2}) reduces to the standard Bekenstein bound (\ref{a1}) in the limit of the HUP regime when $\beta\rightarrow0$. As an interesting result, we see that GUP correction for $\beta<0$ allows the entropy of a thermal system to surpass the Bekenstein bound, while it confines the system to entropies lower than Bekenstein's limit for $\beta>0$. In the following, we first attempt to show that the standard Bekenstein bound (\ref{a1}), with HUP effects, is not satisfied when black hole entropy is described by non-Gaussian statistics contaminated with GUP effects. Then, by using the GUP version of the Bekenstein bound (\ref{a2}), we find how the non-Gaussian statistics indices and GUP parameter $\beta$ are connected in order to fulfill the generalized Bekenstein bound.
\begin{figure*}[!hbtp]
     \centering	\includegraphics[width=0.43\textwidth,keepaspectratio]{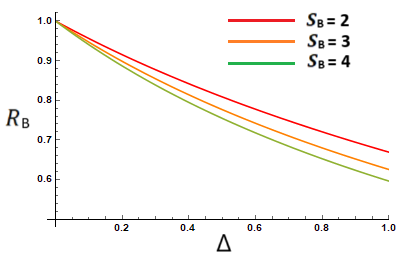}
\includegraphics[width=0.43\textwidth,keepaspectratio]{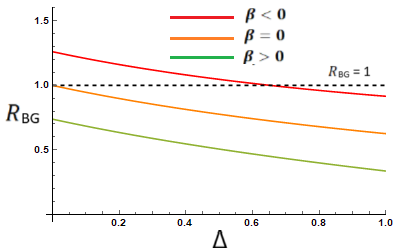}
\caption{\textit{Left}: The parameter $R_{\text{B}}$ versus the Barrow index $\Delta$ for different values of $S_{\text{B}}$. \textit{Right}: The parameter $R_{\text{BG}}$ versus the Barrow index $\Delta$ for different cases of the GUP parameter $\beta$ and $S_{\text{B}}=3$. Both panels are drawn for $k_B=\ell_p=1$.}
\label{fig1}
\end{figure*}

\subsection{Barrow statistics}
Using Koch's snowflake idea \cite{Koch}, Barrow stated that quantum gravitational effects generate a fractal structure on the static, spherically symmetric Schwarzschild black hole surface going to infinity with a finite volume. Thus, the BH entropy can be modified as follows \cite{Barrow:2020tzx}
\begin{equation}
S_{\text{B}}=\bigg(k_B\frac{A}{A_{p}}\bigg)^{1+\frac{\Delta}{2}},
\label{b1}   
\end{equation}
where the subscript $\text{B}$ depicts the Barrow statistics. Here, $A=4\pi R^2$ and $A_{p}=(2\ell_p)^2$, where $\ell_p^2=\hbar G/c^3$, are the surface area of the black hole and the Planck area, respectively. Also, $\Delta$ is the Barrow index referring to the smooth structures for $\Delta=0$ and the most fractal structures for $\Delta=1$. Now, using the surface radius of a Schwarzschild black hole $R=R_{g}=2GM/c^2$, the entropy (\ref{b1}) is rewritten as
\begin{equation}
S_{\text{B}}=\Big(\frac{\pi k_B} {\ell_p^2}\Big)^{1+\frac{\Delta}{2}}R_g^{2+\Delta}.
\label{b2}   
\end{equation}
Due to the quantum effects of gravity applied on black hole surface area of the Schwarzschild black hole, the Bekenstein bound (\ref{a1}) modifies as
\begin{equation}
S_\text{B}\leq\frac{2\pi k_B}{\hbar c}R_g^{1+\frac{\Delta}{2}}E.
\label{b3}    
\end{equation}
By using the formula $E=Mc^2$, then substituting the black hole surface radius from eq.(\ref{b2}), the above bound takes the form
\begin{equation}
S_\text{B}\leq\Big(\frac{\pi k_B}{\ell_p^2}\Big)^{-\frac{\Delta}{4}}S_{\text{B}}^{\frac{4+\Delta}{2(2+\Delta)}}.
\label{b4}    
\end{equation}
It is clear that in the limit of the BH entropy, $\Delta=0$, the above inequality reduces to $S_{\text{BH}}\leq S_{\text{BH}}$, resulting in the Bekenstein bound (\ref{a1}) being satisfied for the Schwarzschild black hole studied in the context of the BH statistics. Moreover, the above inequality tells us that the presence of a non-zero Barrow index, $\Delta\neq0$, will put into question the validity of the modified Bekenstein bound (\ref{b3}). This can be seen in the left panel of Fig.\ref{fig1} where we plot the parameter 
\begin{equation}
R_{\text{B}}\equiv\Big(\frac{\pi k_B}{\ell_p^2}\Big)^{-\frac{\Delta}{4}}S_{\text{B}}^{-\frac{\Delta}{2(2+\Delta)}},
\label{b5}
\end{equation} 
versus the Barrow index $\Delta$ for $S_{\text{B}}>1$. By increasing the value of Barrow's factor $\Delta$, the panel shows more deviation from the consistency condition $R_{\text{B}}=1$. Now, let us pursue the issue from view point of the generalized Bekenstein bound (\ref{a2}) in which HUP effects are replaced by GUP effects. In the context of the Barrow statistics, the generalized Bekenstein bound (\ref{a2}), for $\beta<0$, can be modified as
\begin{equation}
S_{\text{B}}\leq\frac{2\pi k_B}{\hbar c}R_g^{1+\frac{\Delta}{2}}E\Big[1+\frac{|\beta|}{4}\frac{\ell_p^2}{R_g^{2+\Delta}}\Big].
\label{b6}    
\end{equation}
\begin{figure*}[!hbtp]
     \centering	\includegraphics[width=0.43\textwidth,keepaspectratio]{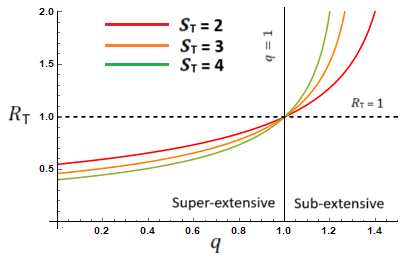}
\includegraphics[width=0.43\textwidth,keepaspectratio]{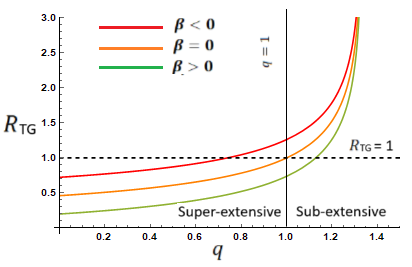}
\caption{\textit{Left}: The parameter $R_{\text{T}}$ versus the Tsallis index $q$ for different values of $S_{\text{T}}$. \textit{Right}: The parameter $R_{\text{TG}}$ versus the Tsallis index $q$ for different cases of the GUP parameter $\beta$ and $S_{\text{T}}=3$. Both panels are drawn for $k_B=1$.}
\label{fig2}
\end{figure*}
Then, after inserting the surface radius from eq.(\ref{b2}), the generalized Bekenstein inequality is given by
\begin{equation}
S_\text{B}\leq\Big(\frac{\pi k_B}{\ell_p^2}\Big)^{-\frac{\Delta}{4}}S_{\text{B}}^{\frac{4+\Delta}{2(2+\Delta)}}+\frac{|\beta|}{4}\ell_p^2\Big(\frac{\pi k_B}{\ell_p^2}\Big)^{1+\frac{\Delta}{4}}S_{\text{B}}^{-\frac{\Delta}{2(2+\Delta)}},
\label{b7}    
\end{equation}
with the corresponding parameter $R_{\text{BG}}$
\begin{equation}
R_\text{BG}\equiv\Big(\frac{\pi k_B}{\ell_p^2}\Big)^{-\frac{\Delta}{4}}S_{\text{B}}^{-\frac{\Delta}{2(2+\Delta)}}+\frac{|\beta|}{4}\ell_p^2\Big(\frac{\pi k_B}{\ell_p^2}\Big)^{1+\frac{\Delta}{4}}S_{\text{B}}^{-\frac{4+3\Delta}{2(2+\Delta)}},
\label{b8}    
\end{equation}
where the subscript $\text{BG}$ depicts the GUP version of the Bekenstein bound in the Barrow statistics. Note that eq.(\ref{b7}) reduces to the satisfied Bekenestin bound in the BH statistics when $\Delta=\beta=0$. Also, by neglecting only the role of the GUP parameter $\beta$, the above result approaches its counterpart in the case of HUP presented in eq.(\ref{b4}). Also, eq.(\ref{b7}) reveals that the presence of the GUP parameter $\beta$ challenges the validity of the Bekenstein bound in the BH regime where we deal with the Schwarzschild black hole with a smooth structure $\Delta=0$. We notice that the deviation from the consistency condition $R_\text{BH}=1$ depends on the sign of the GUP parameter $\beta$ as shown in the right panel of Fig.\ref{fig1} where the parameter $R_\text{BG}$ is drawn versus the Barrow index $\Delta$ for $S_\text{B}=3$ and $\beta=\pm1$. As an important result, the right panel shows that the parameter $R_\text{BG}$ mimics the unit for allowed values of the Barrow index, $0\leq\Delta\leq1$, and a certain value of the Barrow entropy $S_\text{B}$ when $\beta<0$. This means that the generalized Bekenstein bound in Barrow's statistics (\ref{b6}) is satisfied for a possible connection between the statistics factor and the GUP parameter. Therefore, for a certain value of $S_\text{B}$, the parameters $\beta$ and $\Delta$ are connected by
\begin{equation}
R_{\text{BG}}=1\rightarrow|\beta|=\frac{4}{\pi k_{B}}\Big(\frac{\pi k_B}{\ell_p^2}\Big)^{-\frac{\Delta}{2}}S_{\text{B}}\bigg(-1+\Big(\frac{\pi k_B}{\ell_p^2}\Big)^{\frac{\Delta}{4}}S_{\text{B}}^{\frac{\Delta}{2(2+\Delta)}}\bigg).
\label{b9}    
\end{equation}
\subsection{Tsallis statistics}
We start with the non-extensive $q$-generalized statistics based on the non-additive Tsallis entropy \cite{Tsallis:1987eu}
\begin{equation}
S_\text{T}=k_B\frac{1-\sum\limits_{i=1}^Wp_i^q}{q-1},
\label{d1}    
\end{equation}
where the subscript $\text{T}$ depicts the Tsallis statistics. Here, $p_i$ and $W$ are the probability of existence of the system and total number of microstates. Also, the non-extensivity parameter $0\leq q<1$ and $q>1$ correspond to rare and frequent events, respectively. By assuming the same probability for all microstates, Tsallis entropy takes the following from 
\begin{equation}
S_\text{T}=k_B\ln_qW=k_B\frac{W^{1-q}-1}{1-q},
\label{d2}    
\end{equation}
\begin{figure*}[!hbtp]
     \centering	\includegraphics[width=0.43\textwidth,keepaspectratio]{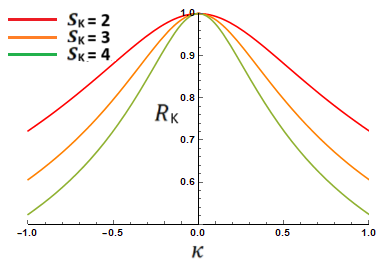}
\includegraphics[width=0.43\textwidth,keepaspectratio]{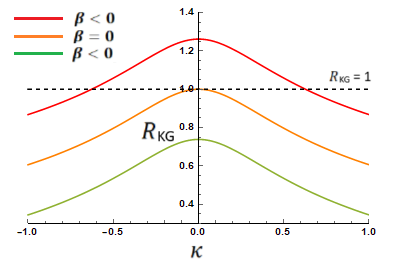}
\caption{\textit{Left}: The parameter $R_{\text{K}}$ versus the Kaniadakis index $\kappa$ for different values of $S_{\text{K}}$. \textit{Right}: The parameter $R_{\text{KG}}$ versus the Kaniadakis index $\kappa$ for different cases of the GUP parameter $\beta$ and $S_{\text{K}}=3$. Both panels are drawn for $k_B=1$.}
\label{fig3}
\end{figure*}
where the Boltzmann-Gibbs entropy, $S=k_B\ln W$, is recovered when $q$ approaches the unit. In the case of the standard Schwarzschild black hole, with $W=e^{\pi R_g^2/\ell_p^2}$, the black hole entropy in the context of Tsallis statistics can be rewritten as
\begin{equation}
S_\text{T}=k_B\frac{e^{(1-q)\frac{\pi R_g^2}{\ell_p^2}}-1}{1-q}.
\label{d3}    
\end{equation}
From the Tsallis statistics perspective, the Bekenstein bound (\ref{a1}) takes the form
\begin{equation}
S_{\text{T}}\leq\frac{2\pi k_B}{\hbar c}R_gE.
\label{d4}    
\end{equation}
From the definition $E=Mc^2$ and then inserting horizon radius from eq.(\ref{d3}), the above bound is given by
\begin{equation}
S_{\text{T}}\leq k_B\frac{\ln\big[\frac{(1-q)}{k_B}S_{\text{T}}+1\big]}{1-q},
\label{d5}    
\end{equation}
which reduces to the satisfied Bekenstein bound in the BH statistics when $q\rightarrow1$. By going beyond the extensive case $q=1$, we find that the modified Bekenstein bound (\ref{d4}) is not satisfied for both super-extensive ($q<1$) and sub-extensive ($q>1$) regions as shown in the left panel of Fig.\ref{fig2}. The panel presents the parameter 
\begin{equation}
R_{\text{T}}\equiv k_B\frac{\ln\big[\frac{(1-q)}{k_B}S_{\text{T}}+1\big]}{(1-q)S_{\text{T}}},
\label{d6}    
\end{equation}
versus the non-extensivity parameter $q$ for $S_\text{T}>1$ in which both non-extensive cases $q<1$ and $q>1$ lead to a deviation from the consistency condition $R_{\text{T}}=1$. The issue can be studied in the presence of the GUP effects by considering the generalized Bekenstein bound (\ref{a2}), for $\beta<0$, in the Tsallis statistics 
\begin{equation}
S_{\text{T}}\leq\frac{2\pi k_B}{\hbar c}R_g E\Big[1+\frac{|\beta|}{4}\Big(\frac{\ell_{pl}}{R_g}\Big)^2\Big].
\label{d7}    
\end{equation}
Then, using the definition $E=Mc^2$ and inserting the black hole radius from eq.(\ref{d3}), the above inequality can be found as
\begin{equation}
S_{\text{T}}\leq k_B\frac{\ln\big[\frac{(1-q)}{k_B}S_{\text{T}}+1\big]}{1-q}+\frac{\pi k_B|\beta|}{4},
\label{d8}    
\end{equation}
with the corresponding parameter $R_{\text{TG}}$
\begin{equation}
R_{\text{TG}}\equiv k_B\frac{\ln\big[\frac{(1-q)}{k_B}S_{\text{T}}+1\big]}{(1-q)S_{\text{T}}}+\frac{\pi k_B|\beta|}{4S_{\text{T}}},
\label{d9}    
\end{equation}
where the subscript $\text{TG}$ depicts the GUP version of the Bekenstein bound in the Tsallis statistics. It is clear that eq.(\ref{d8}) mimics the Bekenstien bound in the BH statistics when $q\rightarrow1$ and $\beta=0$. In the case of $\beta=0$, we reproduce the results of the HUP version of the Bekenstein bound in the Tsallis statistics presented in eq.(\ref{d5}). Moreover, in the limit of the BH statistics, $q\rightarrow1$, the GUP parameter $\beta$ (in both signs) has a destructive role for the consistency condition $R_{\text{BH}}=1$. All above-mentioned results can be seen in the right panel of Fig.\ref{fig2} where the parameter $R_\text{TG}$ is drawn versus the Tsallis index $q$ for $S_\text{T}=3$ and $\beta=\pm1$. Interestingly, the right panel reveals that the parameter $R_\text{TG}$ approaches the unit in super-extensive ($q<1$) and sub-extensive ($q>1$) regions, for a certain value of the Tsallis entropy $S_\text{T}$, when $\beta<0$ and $\beta>0$, respectively. Hence, the generalized Bekenstein bound in the Tsallis statistics (\ref{d7}) is satisfied, for a certain value of $S_\text{T}$, when the parameters $\beta$ and $q$ are connected by
\begin{equation}
R_{\text{TG}}=1\rightarrow\begin{cases}
    \beta=\frac{4}{\pi k_B}\bigg(\frac{k_B}{1-q}\ln\big[1+\frac{(1-q)S_{\text{T}}}{k_B}\big]-S_{\text{T}}\bigg) & \text{for sub-extensive region},\\\\
     |\beta|=\frac{4}{\pi k_B}\bigg(S_{\text{T}}-\frac{k_B}{1-q}\ln\big[1+\frac{(1-q)S_{\text{T}}}{k_B}\big]\bigg) & \text{for super-extensive region}.\\
  \end{cases}
  \label{d10}
\end{equation}

\subsection{Kaniadakis statistics}
Relativistic generalization of Boltzmann–Gibbs entropy navigates us to formulate the $\kappa$-entropy introduced by Kaniadakis as \cite{Kaniadakis:2002zz}
\begin{equation}
S_{\text{K}}=-k_B\sum\limits_{i}^W\frac{p_i^{1+\kappa}-p_i^{1-\kappa}}{2\kappa},
\label{e1}    
\end{equation}
where the subscript K depicts the Kaniadakis statistics. Here, $p_i$ is the system probability, $W$ is number of microstates, and $-1<\kappa<1$ is the Kaniadakis index. Assuming the microcanonical regime, where all microstates have the same probability, Kaniadakis entropy is given by
\begin{equation}
S_{\text{K}}=k_B\ln_\kappa W=k_B\frac{W^{\kappa}-W^{-\kappa}}{2\kappa},
 \label{e2}   
\end{equation}
where the Boltzmann-Gibbs entropy, $S=k_B\ln W$, is recovered when $\kappa\rightarrow0$. For the Schwarzschild black hole, with $W=e^{\pi R_g^2/\ell_p^2}$, the Kaniadakis entropy (\ref{e2}) takes the following form
\begin{equation}
S_{\text{K}}=k_B\frac{e^{\frac{\pi\kappa R_g^2}{\ell_p^2}}-e^{-\frac{\pi\kappa R_g^2}{\ell_p^2}}}{2\kappa}.
\label{e3}    
\end{equation}
On the other hand, the Bekenstein conjecture bound (\ref{a1}) in the context of the Kaniadakis statistics modifies as
\begin{equation}
S_{\text{K}}\leq\frac{2\pi k_B}{\hbar c}R_gE.
\label{e4}    
\end{equation}
Then, by using the definition $E=Mc^2$ and substituting the black hole radius from eq.(\ref{e3}), the above inequality can be rewritten as
\begin{equation}
S_{\text{K}}\leq\frac{k_B}{\kappa}\Bigg(2\pi iC+\ln\Big[\frac{\kappa S_\text{K}\pm\sqrt{k_B^2+\kappa^2S_\text{K}^2}}{k_B}\Big]\Bigg),
\label{e5}    
\end{equation}
where the constant $C$ is an integer so that we fix it as zero in the rest of the paper. Note that by choosing the positive sign and sending the Kaniadakis index $\kappa$ to zero, the above bound reduces to the satisfied Bekenstien bound in the BH statistics. Also, eq.(\ref{e5}) discloses that the modified Bekenstein bound (\ref{e4}) can not be fulfilled when we deal with a non-zero Kaniadakis factor $\kappa$. This result is shown in the left panel of Fig.\ref{fig3}, where we draw the parameter 
\begin{equation}
R_{\text{K}}\equiv\frac{k_B}{\kappa S_{\text{K}}}\ln\Big[\frac{\kappa S_\text{K}+\sqrt{k_B^2+\kappa^2S_\text{K}^2}}{k_B}\Big],
\label{e6}    
\end{equation}
versus the Kaniadakis index $\kappa$ for $S_{\text{K}}>1$ in which the consistency condition $R_{\text{K}}=1$ is perturbed for both positive and negative cases of the Kaniadakis index $\kappa$. To study the issue in the GUP regime, we rewrite the generalized Bekenstein bound (\ref{a2}) in the Kaniadakis statistics, for $\beta<0$, as follows
\begin{equation}
S_{\text{K}}\leq\frac{2\pi k_B}{\hbar c}R_g E\Big[1+\frac{|\beta|}{4}\Big(\frac{\ell_{pl}}{R_g}\Big)^2\Big].
\label{e7}    
\end{equation}
By substituting $E=Mc^2$ and the black hole radius from eq.(\ref{e3}), the above inequality can be found as 
\begin{equation}
S_{\text{K}}\leq\frac{k_B}{\kappa}\ln\Big[\frac{\kappa S_\text{K}+\sqrt{k_B^2+\kappa^2S_\text{K}^2}}{k_B}\Big]+\frac{\pi k_B|\beta|}{4},
\label{e8}    
\end{equation}
with the corresponding parameter $R_{\text{KG}}$
\begin{equation}
R_{\text{KG}}\equiv\frac{k_B}{\kappa S_{\text{K}}}\ln\Big[\frac{\kappa S_\text{K}+\sqrt{k_B^2+\kappa^2S_\text{K}^2}}{k_B}\Big]+\frac{\pi k_B|\beta|}{4S_{\text{K}}},
\label{e9}    
\end{equation}
where the subscript $\text{KG}$ depicts the GUP version of the Bekenstein bound in the Kaniadakis statistics. It is easy to see that the satisfied Bekenstein bound in the BH statistics is recovered from the bound (\ref{e8}) when $\kappa\rightarrow0$ and $\beta=0$. Also, the bound (\ref{e8}) reduces to the case of the HUP (\ref{e5}) by removing the role of the GUP parameter $\beta$. Analogous to the previous statistics, the GUP parameter $\beta$ disturbs the satisfied Bekenstein bound in the BH realm where the Kaniadakis index $\kappa$ goes to zero. The right panel of Fig.\ref{fig3} shows the behaviour of the parameter $R_{\text{KG}}$ versus the Kaniadakis index $\kappa$ for $S_{\text{K}}=3$ and $\beta=\pm1$ in which the consistency condition $R_{\text{BH}}=1$ is deviated for both signs of the GUP parameter $\beta$. Besides the mentioned results, the panel tells us that the parameter $R_{\text{KG}}$ approaches the unit for both signs of the Kaniadakis index $\kappa$, for a certain value of the Kaniadakis entropy  $S_{\text{K}}$, when $\beta<0$. Consequently, the generalized Bekenstein bound in the context of the Kaniadakis statistics (\ref{e7}) is valid for a certain value of $S_{\text{K}}$ and the following connection between $\beta$ and $\kappa$
\begin{equation}
R_{\text{KG}}=1\rightarrow|\beta|=\frac{4}{\pi k_B}\bigg(S_{\text{K}}-\frac{k_B}{\kappa}\ln\big[\frac{\kappa S_\text{K}+\sqrt{k_B^2+\kappa^2S_\text{K}^2}}{k_B}\big]\bigg).
\label{e10}    
\end{equation}
\section{Conclusions}\label{s3}
According to the second law of thermodynamics, the entropy of a closed physical system can be increased until it approaches a maximum point. Hence, universal entropy bounds have been proposed to confine the growth of the entropy in thermal systems down to the quantum level. Inspired by the black hole discussions, the Bekenstein bound conjecture was proposed to restrict the upper limit of the entropy of both gravitational and non-gravitational phenomena. In the present work, we first studied the Bekenstein bound on the entropy of a Schwarzschild black hole described by some non-Gaussian statistics such as Barrow, Tsallis, and Kaniadakis. We realized that Bekenstein's bound is violated in the presence of non-vanished indices $\Delta$, $q$ and $\kappa$ related to the mentioned statistics. Then, by considering the GUP version of the Bekenstein bound, we found that the generalized bound is satisfied in the context of the above-mentioned statistics due to a relation between the entropies indices and the GUP parameter $\beta$. 

In summary, the standard Bekenstein bound is \textit{partially} satisfied in non-Gaussian statistics since its upper limit (\ie $R_\text{B}=1$ for Barrow, $R_\text{T}=1$ for Tsallis and $R_\text{K}=1$ for Kaniadakis) is not achievable, while the GUP version of the Bekenstein bound provides a framework for a \textit{fully} satisfied black hole bound in non-Gaussian statistics due to the increase in entropy, which usually happens for $\beta<0$.

\bibliographystyle{ieeetr}
\bibliography{biblo}
\end{document}